\documentclass[11pt,twoside]{article}
\usepackage{asp2010}

\usepackage{epsf}

\resetcounters

\begin{document}

\title{MESA and NuGrid Simulations of Classical Nova Outbursts and Nucleosynthesis}
\author{Pavel Denissenkov,$^1$ Falk Herwig,$^1$ Marco Pignatari,$^2$ and James Truran$^3$
\affil{$^1$Department of Physics \& Astronomy, University of Victoria, P.O.~Box 3055, Victoria, B.C., V8W~3P6, Canada}
\affil{$^2$Department of Physics, University of Basel, Klingelbergstrasse 82, CH-4056 Basel, Switzerland}
\affil{$^3$Department of Astronomy and Astrophysics, and Enrico Fermi Institute, University of Chicago, 5640 South Ellis Avenue Chicago, Chicago, IL 60637 USA}}

\begin{abstract}
Classical novae are the results of surface thermonuclear explosions of hydrogen
accreted by white dwarfs (WDs) from their low-mass main-sequence or red-giant binary companions.
Chemical composition analysis of their ejecta shows that nova outbursts
occur on both carbon-oxygen (CO) and more massive oxygen-neon (ONe) WDs, and
that there is cross-boundary mixing between the accreted envelope and underlying WD.
We demonstrate that the state-of-the-art stellar evolution code MESA and
post-processing nucleosynthesis tools of NuGrid can successfully be used
for modeling of CO and ONe nova outbursts and nucleosynthesis.
The convective boundary mixing (CBM) in our 1D numerical simulations is implemented using
a diffusion coefficient that is exponentially decreasing with a distance below the bottom of
the convective envelope. We show that this prescription produces maximum temperature evolution
profiles and nucleosynthesis yields in good agreement with those obtained using the commonly adopted
1D nova model in which the CBM is mimicked by assuming that the accreted envelope has been
pre-mixed with WD's material. In a previous paper, we have found that $^3$He can be produced {\it in situ}
in solar-composition envelopes accreted with slow rates ($\dot{M} < 10^{-10}\,M_\odot/\mbox{yr}$) by cold ($T_{\rm WD} < 10^7$ K)
CO WDs, and that convection is triggered by $^3$He burning before the nova outburst in this case. Here, we confirm
this result for ONe novae. Additionally, we find that the interplay between the $^3$He
production and destruction in the solar-composition envelope accreted with an intermediate rate, e.g.
$\dot{M} = 10^{-10}\,M_\odot/\mbox{yr}$, by the $1.15\,M_\odot$ ONe WD with a relatively high initial central temperature,
e.g. $T_{\rm WD} = 15\times 10^6$ K,
leads to the formation of a thick radiative buffer zone that separates the bottom of the convective
envelope from the WD surface.
\end{abstract}

\section{Introduction}
MESA is a collection of Fortran-95 {\bf M}odules for {\bf E}xperiments in {\bf S}tellar {\bf A}strophysics.
Its main module {\tt star} can be used for 1D stellar evolution simulations of almost any kind (numerous examples
are provided by \cite{paxton:11,paxton:13} and on the MESA website {\tt http://mesa.sourceforge.net}).
For instance, {\tt star} can compute without any interruption the evolution of a solar-type star from the pre-MS
phase through the He-core flash and thermal pulses on the asymptotic giant branch towards the WD cooling.
Only a few codes can do such computations.
Other MESA modules provide {\tt star} with state-of-the-art numerical algorithms, e.g. for adaptive mesh refinement and
timestep control, atmospheric boundary conditions, and modern input physics (opacities, equation of state, nuclear
reaction rates, etc.) 

The nucleosynthesis calculations presented in this work
are made by the post-processing code PPN \citep[developed inside the NuGrid research
framework, ][]{herwig:08}, where the input data of
stellar structure (e.g., from MESA models) are processed using a
comprehensive list of nuclear species and reaction rates.
The network can include more than 5000 species, between H and Bi, and more than 50000
nuclear reactions. A self-controlled dynamical network defines the actual numbers of species and reactions
considered in calculations, based on the strength of nucleosynthesis fluxes.
For the complete post-processing of the full MESA nova models, we use the multi-zone parallel frame of PPN (MPPNP),
whereas for the simulations based on a single-zone trajectory we use the serial frame (SPPN).
The two frames use the same nuclear physics library and the same package to solve the nucleosynthesis equations
(for details, go to {\tt http://www.nugridstars.org}).

We have prepared shell scripts, template MESA inlist files, and a large number of initial WD models aiming to
combine MESA and NuGrid into an easy-to-use Nova Framework.
This new research tool can model
nova outbursts and nucleosynthesis occurring on CO and ONe WDs using up-to-date input physics and a specified
nuclear network for a relatively large number (up to a few thousands) of mass zones covering both the WD and its accreted
envelope. Nova simulation results can be analyzed using animations and a variety of plots produced with NuGrid's Python
visualization scripts. This article reports the new results obtained with the Nova Framework and demonstrates some of 
its capabilities.

\section{One-Dimensional Nova Models With Convective Boundary Mixing (CBM)}

\begin{figure}[!ht]
\plottwo{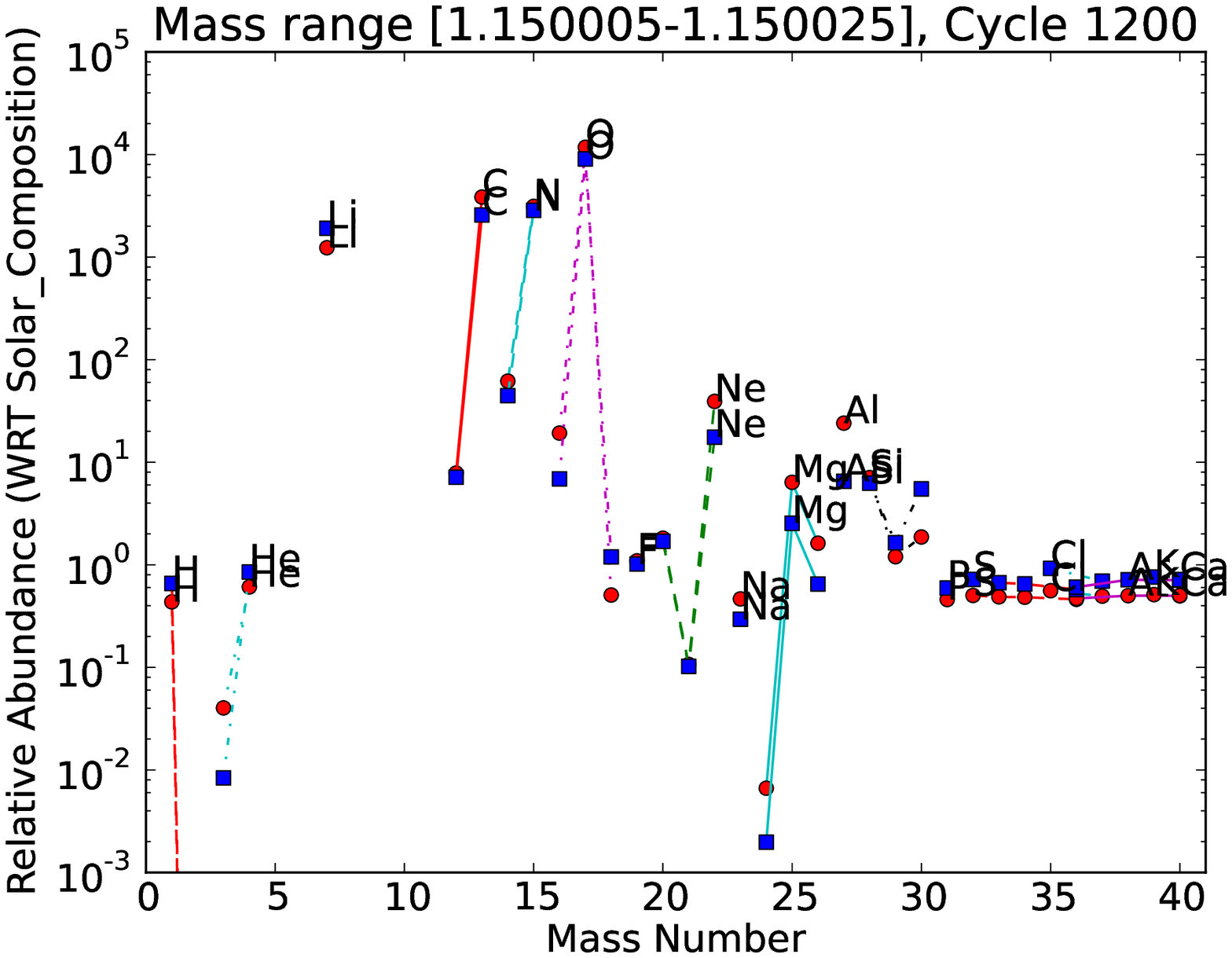}{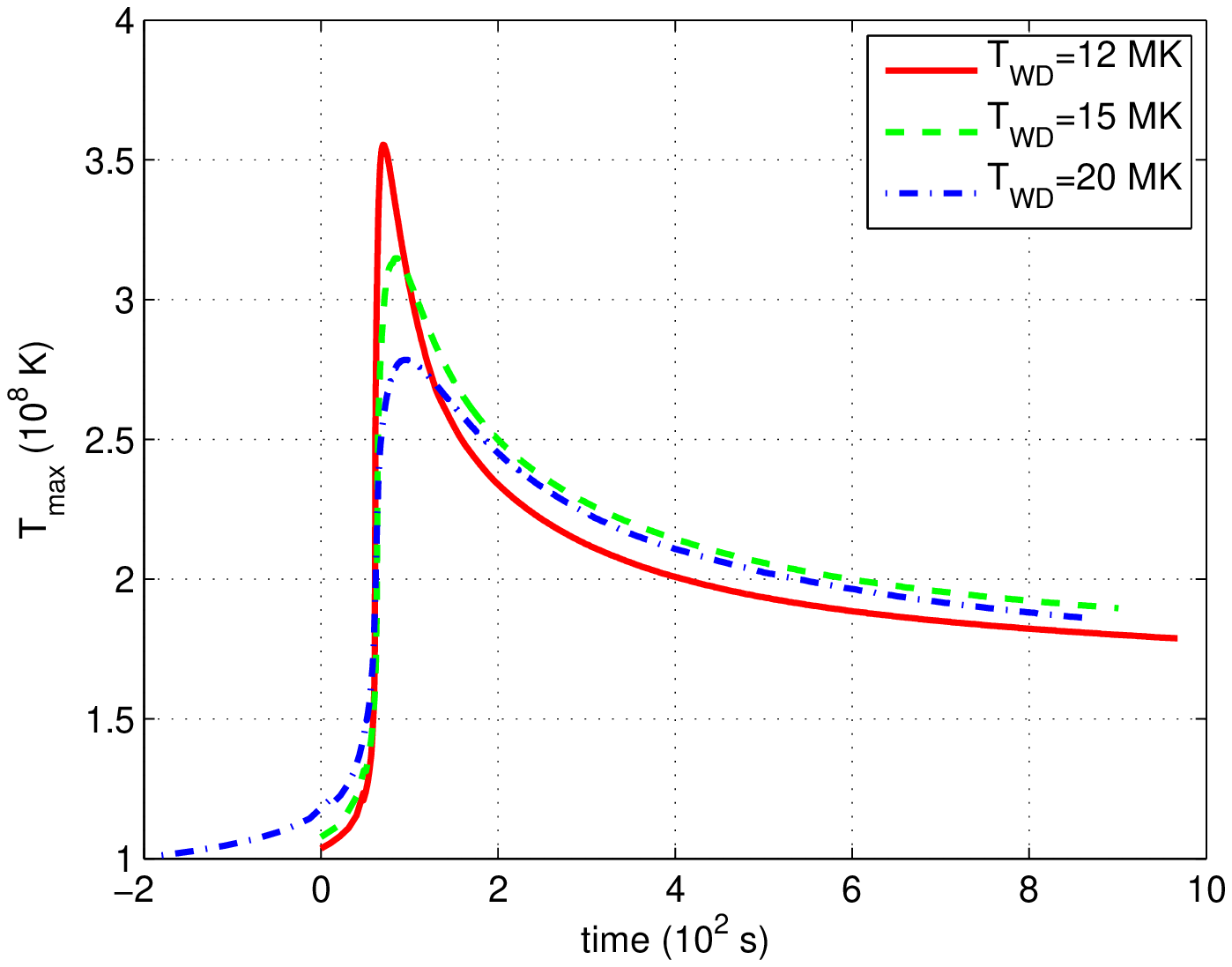}
\caption{Left panel: A comparison of final solar-scaled abundances of stable isotopes from two of our
         $1.15\,M_\odot$ CO nova simulations with $T_{\rm WD} = 12$ MK and $\dot{M} = 2\times 10^{-10}\,M_\odot/\mbox{yr}$.
         The red circles are the results obtained with the 50\% pre-mixed
         initial abundances in the accreted envelope, while the blue squares represent the case
         with the solar-composition accreted material and convective boundary mixing (CBM) modeled using the same method
         and CBM parameter $f_{\rm nova} = 0.004$ that we used in the $1.2\,M_\odot$ CO nova model in Paper~I.
         Right panel: $T_{\rm max}$-trajectories from three of our $1.3\,M_\odot$ 50\% pre-mixed ONe nova models 
         computed for the same accretion rate, $\dot{M} = 2\times 10^{-10}\,M_\odot/\mbox{yr}$,
         but different WD's initial central temperatures.
         The initial isotope abundances are those of the Barcelona group \citep{jose:99,jose:01}.
         }
\label{fig:f1}
\end{figure}

To be consistent with observations, a realistic nova model should assume that the accreted material has solar composition
(for the solar metallicity) and that it is mixed with WD's material before or during its thermonuclear runaway.
Recent two- and three-dimensional nuclear-hydrodynamic simulations of a nova outburst have shown that a
possible mechanism of this mixing are the hydrodynamic instabilities and shear-flow turbulence induced by steep horizontal
velocity gradients at the bottom of the convection zone triggered by the runaway
\citep{casanova:10,casanova:11a,casanova:11b}. These hydrodynamic processes associated with the
convective boundary lead to convective boundary mixing (CBM) at the
base of the accreted envelope into the outer layers of the WD.
As a result, CO-rich (or ONe-rich) material is dredged-up during the runaway.
In our 1D MESA simulations of nova outbursts, we use a simple CBM model that treats the time-dependent mixing as a
diffusion process. The model approximates the rate of mixing by an exponentially decreasing function of a distance from the formal
convective boundary,
\begin{eqnarray} D_{\rm CBM} = D_0\exp\left(-\frac{2|r-r_0|}{fH_P}\right),
\label{eq:DCBM}
\end{eqnarray} where $H_P$ is the pressure scale height, and $D_0$ is
a diffusion coefficient, calculated using a mixing-length theory, that
describes convective mixing at the radius $r_0$ close to the
boundary. In this model $f$ is a free parameter, for which we use the same value $f = f_\mathrm{nova} = 0.004$
that we used in our CO nova simulations in Paper~I \citep{denissenkov:13}. 
For a physical motivation to model the CBM with the prescription
(\ref{eq:DCBM}), see Paper~I. Our MESA nova models with the CBM have been post-processed with the MPPNP code,
and the resulting
final abundances have been compared with those obtained for the corresponding nova models without CBM but with the 50\% pre-mixed
accreted envelopes. The comparison between the final abundances shows a very good agreement
for both the CO and ONe nova models (the left panel of Fig.~\ref{fig:f1} and Fig. 1 in \citealt{denissenkov:12};
we disavow our conclusion made in that paper 
that it was not true for CO novae).  This gives a support to the widely-adopted 1D nova model
in which the CBM is mimicked by assuming that the accreted envelope
has been pre-mixed with WD's material.

\begin{figure}[!ht]
\plottwo{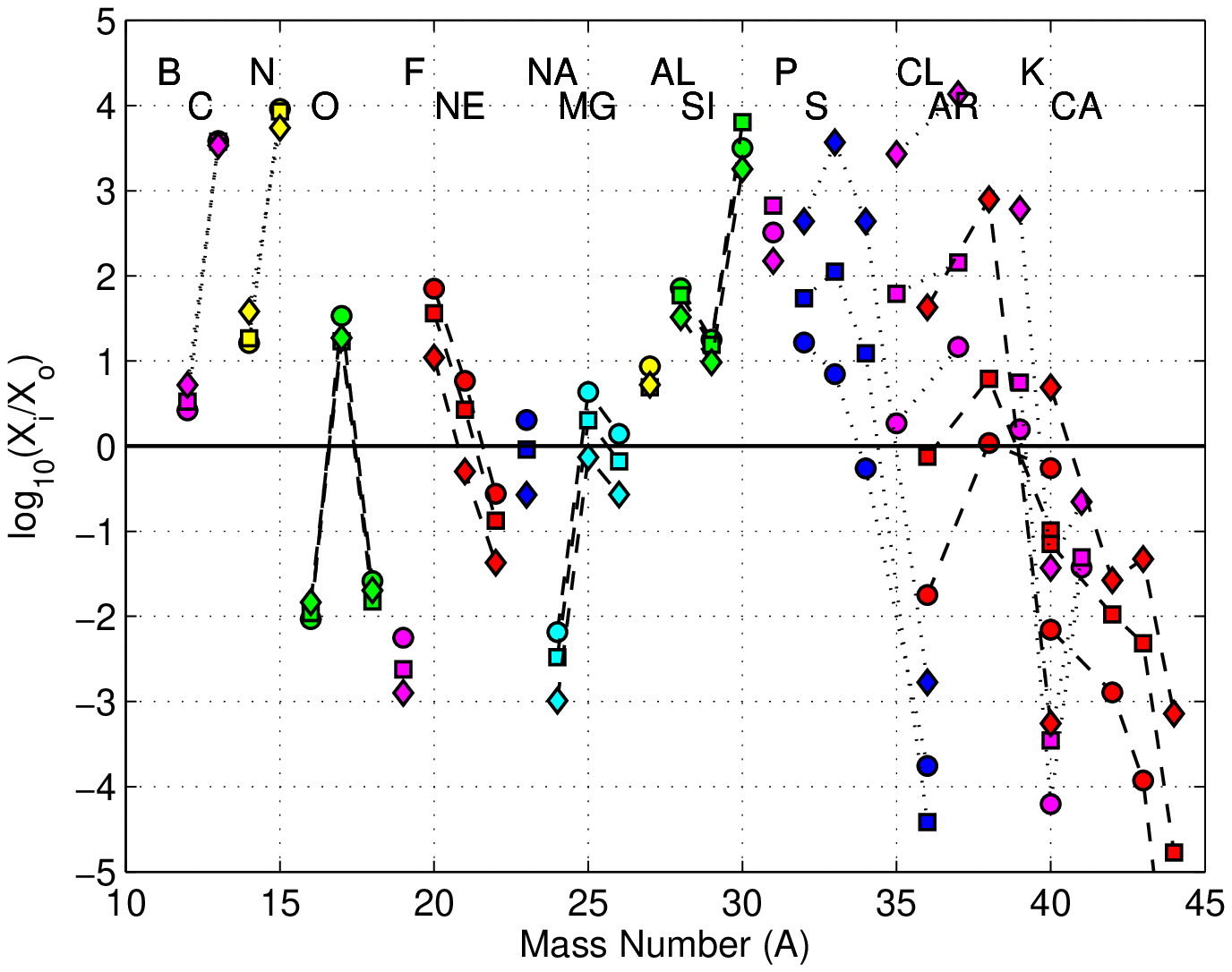}{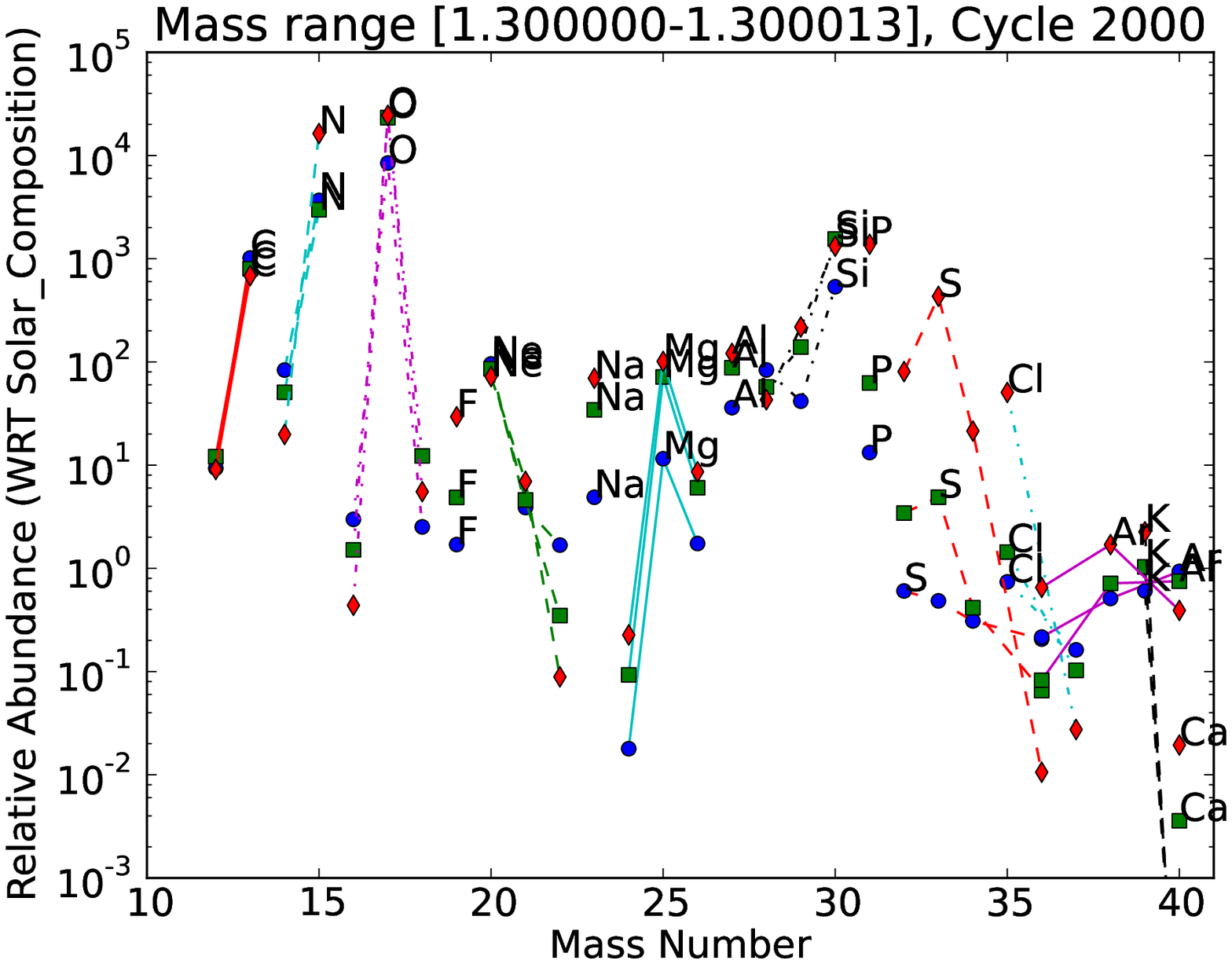}
\caption{Left panel: the final solar-scaled abundances of stable isotopes calculated with the SPPN code
         using the trajectories from the right panel of Fig.~\ref{fig:f1} that correspond to the three different
         WD's initial central temperatures: $T_\mathrm{WD} = 12$ MK (diamonds), 15 MK (squares), and 20 MK (circles).
         Right panel: the same final abundances, but calculated with the MPPNP code using the corresponding full MESA models
         and averaged over the indicated mass range in the expanding nova envelope. 
         }
\label{fig:f2}
\end{figure}

\section{Comparison of the Results of SPPN and MPPNP Simulations}

Single-zone post-processing calculations of nova nucleosynthesis with the NuGrid code SPPN are done for a mass zone
in a nova envelope where the temperature profile has its maximum, $T=T_\mathrm{max}$ 
(usually, near the core-envelope interface). For this,
the temporal variations of $T_\mathrm{max}$ and its corresponding density, called ``trajectories'', are extracted from MESA models
(e.g., the right panel of Fig.~\ref{fig:f1}). The SPPN code runs much faster than MPPNP, therefore it can be used for 
a comprehensive numerical analysis of parameter space, when the relevant isotope abundances and reaction rates are varied
within their observationally and experimentally constrained limits. 
A comparison between the final abundance distributions in Fig.~\ref{fig:f2}
shows a qualitative agreement between the results
from the single-zone $T_\mathrm{max}$-trajectories (the left panel) and from their corresponding multi-zone 
complete models (the right panel).
Indeed, the abundance patterns for different groups of isotopes look similar.
The main quantitative difference is the more efficient production toward heavier species
in the first case, where O is more depleted, while Ar and K isotopes\footnote{The Barcelona composition specifies
the initial abundances only
for isotopes lighter than Ca.} are more efficiently made.
This difference is caused by the fact that in the multi-zone post-processing simulations
convective mixing reduces an ``average temperature'' at which the nucleosynthesis occurs and it also
constantly replenishes the hydrogen fuel burnt at the base of the envelope by bringing it from its outer parts,
where $T\ll T_\mathrm{max}$.
In general, the results in Fig.~\ref{fig:f2} show how the increase
of the $T_\mathrm{max}$ peak value (the right panel of Fig.~\ref{fig:f1}) changes the relative abundance distribution.
We conclude that SPPN nucleosynthesis simulations with nova $T_\mathrm{max}$-trajectories provide
a proper qualitative indication of the behavior of their corresponding complete MPPNP simulations.
However, they can be used only as a diagnostic for nova nucleosynthesis, e.g.
in reaction rate sensitivity studies.

\section{Effects Caused by $^3$He Burning}

In Paper~I, we have found, for the first time, 
that in the extreme case of a very cold CO WD, e.g. with $T_\mathrm{WD} = 7$ MK,
accreting solar-composition material with a very low rate, say $\dot{M} = 10^{-11} M_\odot/\mathrm{yr}$, the incomplete
pp~I chain reactions lead to the {\em in situ} synthesis of $^3$He in a slope adjacent to the base of the accreted
envelope and that the ignition of this $^3$He triggers convection before the major nova outburst. The $^3$He burning
continues at a relatively low temperature, $T\approx 30$ MK, approximately until its abundance is reduced below the solar value,
only after that the major nova outburst ensues triggered by the reaction $^{12}$C(p,$\gamma)^{13}$N. Although this is
an interesting variation of the nova scenario,\footnote{Initially, $^3$He was considered as the most likely isotope
to trigger a nova outburst by \cite{schatzman:51}.} because the CO enrichment of the accreted envelope is produced by
the $^3$He-driven CBM in this case, its observational frequency is expected to be very low (Paper~I).
Later, we have confirmed this result for ONe novae \citep[Fig. 2 in ][]{denissenkov:12}. Furthermore, we have found that
in the $1.15\,M_\odot$ ONe WD with a relatively high initial central temperature, $T_{\rm WD} = 15$ MK, accreting
the solar-composition material with the intermediate rate, $\dot{M} = 10^{-10}\,M_\odot/\mbox{yr}$,
the interplay between the $^3$He production and destruction in the vicinity of the base of the accreted envelope
first leads to a shift of $T_\mathrm{max}$ away from the core-envelope interface followed by the formation of a thick
radiative buffer zone that separates the bottom of the convective envelope from the WD surface \citep[Fig. 3 in ][]{denissenkov:12}.
This result is obtained only for the case when an ONe WD accretes solar-composition material and it almost disappears
in the models with the pre-mixed accreted envelopes. Given that the formation of the radiative buffer zone is revealed
in the computations with the more realistic nova model parameters, such cases can probably be observed.
We defer a more detailed study of this peculiar case and its possible consequences to a future work.
In particular, it would be interesting
to see if the CBM can cross the buffer zone and reach the WD.

\section{Conclusion}

We have created the Nova Framework that allows to simulate the accretion of H-rich material
onto a white dwarf (WD) leading, for a suitable set of initial parameters, to a nova outburst,
and to post-process its accompanying nucleosynthesis.
The Nova Framework combines the state-of-the-art stellar evolution code MESA and post-processing nucleosynthesis
tools of NuGrid. It includes a number of CO and ONe WD models with different
masses and central temperatures (luminosities) that can be used in simulations of
nova outbursts. The use of the Nova Framework is facilitated by a number of shell scripts
that do the routine job necessary to coordinate the operation of the MESA and NuGrid codes.

Within the Nova Framework project, we have provided a large set of nucleosynthesis
calculations for nearly 50 CO and ONe nova models at the solar
metallicity ($Z = 0.02$) so far. To verify our calculations, we have compared their results with
those published in the literature for similar nova models. The comparison shows
a very good qualitative agreement.

\acknowledgements This research has been supported by the National
Science Foundation under grants PHY 11-25915 and AST 11-09174. This
project was also supported by JINA (NSF grant PHY 08-22648) and
TRIUMF. FH acknowledges funding from NSERC through a Discovery
Grant. MP also thanks the support from Ambizione grant of the SNSF
(Switzerland), and from EuroGENESIS.

\bibliographystyle{asp2010}
\bibliography{my_asp}

\end{document}